\documentclass[12pt,preprint]{aastex}

\usepackage{natbib}
\usepackage{amsbsy}

\newcommand{\bfvarepsilon}{\boldsymbol{\varepsilon}}
\newcommand{\bfomega}{\boldsymbol{\omega}}

\newcommand{\ddeg}{\hbox{.\hskip-2pt $^\circ$}}

\newcommand{\kms}{\,km\,s$^{-1}$}
\newcommand{\masyr}{\,mas\,yr$^{-1}$}
\newcommand{\muas}{\,$\mu$as}
\newcommand{\muasyr}{\,$\mu$as\,yr$^{-1}$}

\shorttitle{Tilting of the Milky Way disk}
\shortauthors{Perryman et al.}

\begin{document}

\title{The Gaia inertial reference frame \\and the tilting of the Milky Way disk}

\author{Michael Perryman\altaffilmark{1,2} and David N. Spergel}
\affil{Department of Astrophysical Sciences, Peyton Hall, Princeton, NJ~08544}

\and 

\author{Lennart Lindegren}
\affil{Lund Observatory, Lund, Box~43, 22100 Sweden}

\altaffiltext{1}{Bohdan Paczy{\' n}ski Visiting Fellow}
\altaffiltext{2}{Adjunct Professor, School of Physics, University College Dublin, Ireland}

\begin{abstract} 
While the precise relationship between the Milky Way disk and the symmetry planes of the dark matter halo remains somewhat uncertain, a time-varying disk orientation with respect to an inertial reference frame seems probable. Hierarchical structure formation models predict that the dark matter halo is triaxial and tumbles with a characteristic rate of $\sim$2\,rad/Hubble time ($\sim$30\muasyr). These models also predict a time-dependent accretion of gas, such that the angular momentum vector of the disk should be misaligned with that of the halo. These effects, as well as tidal effects of the LMC, will result in the rotation of the angular momentum vector of the disk population with respect to the quasar reference frame. We assess the accuracy with which the positions and proper motions from Gaia can be referred to a kinematically non-rotating system, and show that the spin vector of the transformation from any rigid self-consistent catalog frame to the quasi-inertial system defined by quasars should be defined to better than~1\muasyr. Determination of this inertial frame by Gaia will reveal any signature of the disk orientation varying with time, improve models of the potential and dynamics of the Milky Way, test theories of gravity, and provide new insights into the orbital evolution of the Sagittarius dwarf galaxy and the Magellanic Clouds.
\end{abstract}



\section{Introduction}

The Gaia space astrometry mission will make precision measurements of the positions and motions of both Galactic stars and distant quasars. Like Hipparcos before it, Gaia will utilize a small number of key measurement principles (observations above the atmosphere, two widely-separated viewing directions, and a uniform `revolving scanning' of the celestial sphere) to create catalogs of star positions, proper motions, and parallaxes of state-of-the-art accuracies \citep{2001A&A...369..339P,2008IAUS..248..217L}. Crucially, both generate {\it absolute\/} trigonometric parallaxes, rather than the relative parallaxes accessible to narrow-field astrometric measurements from the ground. In both cases, the observations are effectively reduced to an internally consistent and extremely `rigid' catalog of positions and proper motions, but whose frame orientation and angular rate of change (spin) are essentially arbitrary, since the measured arc lengths between objects are invariant to frame rotation. Placing both positions and proper motions on an inertial system corresponds to determining these 6~degrees of freedom (3~orientation and 3~spin components). They were derived after catalog completion for Hipparcos, and will be derived as a by-product of the observations/data reductions in the case of Gaia. 

Will Gaia measure the same fundamental plane for quasars and for Galactic stars? Simulations of galaxy formation \citep{2004ApJ...616...27B,2007MNRAS.380..657B} predict that most galaxy halos tumble with a characteristic rotation rate of $\sim 2$\,rad/Hubble time. Both analytical arguments \citep{1995MNRAS.275..897N} and numerical simulations \citep{1995ApJ...442..492D} suggest that dynamical friction in the inner regions of galaxies should tightly couple the inner disk to the halo, at least to $\sim R_{\rm vir}$ \citep{2004ApJ...616...27B}, corresponding roughly to the region dominated by the baryons. Thus, if the angular momentum vectors of the inner disk and halo remain aligned \citep{2007MNRAS.374...16L}, we would expect that the fundamental plane defined by the Galactic stars will rotate at a rate of $\sim 30$\muasyr. 

Even in the absence of a tumbling halo contribution, the disk orientation is expected to vary with time, due to a combination of the infall of misaligned gas \citep{2006MNRAS.370....2S}, the interaction of the infalling gas with the halo \citep{2010MNRAS.408..783R}, and the effect of the LMC \citep{2006ApJ...641L..33W}.  Simulations show the inner disk and the outer halo often decouple \citep{2010MNRAS.408..783R}, with average misalignments of $30^\circ-40^\circ$ \citep{2009MNRAS.400...43C,2010MNRAS.404.1137B,2010MNRAS.405..274H}, such that the reference frame defined by the inner and outer disk stars may differ.  \citet{2013MNRAS.434.2971D} suggest that the observations of the Sagittarius stream may be better fit by models where the outer disk is not aligned with the principal plane of the dark matter halo. If any of these effects apply, Gaia may measure a different fundamental plane for quasars, inner disk stars and outer disk stars.

For Hipparcos, typical positions and annual proper motions were of order 1\,mas (milli-arcsec), and the reference frame link was determined with an accuracy of about 0.6\,mas in the three orientation components, and 0.25\masyr\ in the three spin components, determined by a variety of different link methodologies (described further below). Gaia will achieve accuracies of some 10\muas\ (micro-arcsec) in positions and annual proper motions for bright stars ($V\sim10$), degrading to around 25\muas\ at $V=15$, and to around 0.3\,mas (300\muas) at $V=20$ \citep{2008IAUS..248..217L}. 

This paper addresses the accuracy with which Gaia can detect the rotation of the angular momentum vector defined by disk stars relative to the inertial frame defined by quasars. We will show that the reference frame link should be determined to better than 1\muasyr\ in spin. Being significantly smaller than (for example) dynamical effects driven by a tumbling halo, we argue that accurately linking the Gaia catalog to an inertial reference system will, in addition to its expected impacts in many other fields of stellar kinematics, deepen our understanding of the larger scale dynamics and history of the Milky Way.

\section{Reference systems and reference frames}

The IAU Working Group on Reference Frames and Reference Systems emphasizes the distinction between the theoretical construct of a celestial reference `system', and its practical materialization, referred to as a reference `frame', via a set of fiducial astronomical sources, whether at optical, radio or other wavelengths.

Historically, celestial reference {\it systems\/} were referred to the position of the Earth's equator and equinox at some specified epoch. Thus the reference {\it system\/} B1950 specified positions with respect to (an outward extension of) the Earth's equator, and to the equinox location at epoch B1950.0. It was materialized by the FK4 reference {\it frame\/} comprising positions and proper motions of the 1535 stars of the fundamental catalog~FK4. B1950/FK4 was later superseded by J2000/FK5, viz.\ the {\it reference system\/} J2000 (i.e.\ referred to the equator/equinox at epoch J2000), materialized by the FK5 reference {\it frame}, comprising improved positions/proper motions of the same 1535~primary reference stars, along with some 3000 others (the FK5 extension).

The International Celestial Reference System (ICRS) superseded the J2000 equator/equinox-based system, with the goal of placing positions and proper motions of celestial objects directly on an (extragalactic-based) inertial reference system. It was materialized by the International Celestial Reference Frame (ICRF), initially consisting of positions of 212 extragalactic radio sources, observed at 2.3 and 8.4\,GHz by Mark~III VLBI through the middle of 1995, and with rms positional uncertainty between 100--500\muas\ \citep{1998AJ....116..516M}. The IAU adopted the ICRF as the fundamental celestial reference frame, superseding the FK5 optical frame as of 1998 January~1. More recently, the ICRF2 has been extended to include positions of 3414~extragalactic radio sources observed by VLBI over 30~years, with an improved noise floor of $\sim$40\muas, and an improved axis stability of $\sim$10\muas\ \citep{2009ITN....35....1M}.

\subsection{The Hipparcos reference frame}

Finalising the Hipparcos catalog included adjustments in both orientation and spin components such that the Hipparcos reference frame coincided with the ICRF, as already established in the radio. Following publication in 1997, the IAU adopted the Hipparcos catalog as the {\it optical\/} materialization of the ICRS. With a completeness limit of 7.3--9.0\,mag, and a faint star limit of $V\sim12$, Hipparcos included just one extragalactic object, the quasar 3C~273, and that with rather poor positional precision reflecting its faint magnitude. Accordingly, a number of different approaches were pursued, in parallel, to establish the 6~link parameters \citep{1997A&A...323..620K}. These were:
(i)~interferometric observations of radio stars by VLBI, MERLIN and VLA; 
(ii)~observations of quasars relative to Hipparcos stars via CCDs, photographic plates, and HST; 
(iii)~photographic programmes to determine stellar proper motions with respect to extragalactic objects; 
and (iv)~comparison of Earth orientation parameters obtained by VLBI and by ground-based optical observations of Hipparcos stars. 
The various techniques generally agreed to within 10\,mas in the orientation components, and to within 1\,mas\,yr$^{-1}$ in spin components. Weighted mean values were adopted for the definition of the system of positions and proper motions. As a result, the coordinate axes defined by the published catalog (at catalog mid-epoch, J1991.25) were considered aligned to the extragalactic radio frame with rms uncertainties estimated to be 0.6\,mas in the three components of the orientation vector, $\bfvarepsilon$, and 0.25\masyr\ in the three components of the spin vector, $\bfomega$ (we adopt Galactic coordinates, with $\omega_1$ towards the Galactic center, $\omega_2$ in the direction of Galactic rotation, and $\omega_3$ towards the Galactic pole).

Numerous subsequent studies, including those with a longer temporal baseline, have largely confirmed these values. Some have hinted at slightly larger spin components in $\omega_3$ \citep[e.g.,][]{2004AstL...30..848B,2011MNRAS.416..403F,2013MNRAS.430.2797A}, although this is the most sensitive to the effects of (differential) Galactic rotation.
As an independent verification of the link, the kinematic bulk motion of Galactic disk stars within the adopted reference frame reveals no unexpected rotational component about axes in the plane of the Galaxy ($\omega_1,\omega_2$), although such bulk motions, even if present, would not in themselves invalidate the accuracy of the claimed link.

Subsequent kinematic investigations of the Hipparcos proper motions within $\sim$3\,kpc have shown warp-like structures but of confusing and conflicting form \citep{1998Natur.392..471S,2000A&A...354...67D}. Warps are a common feature of a large fraction of spiral galaxies \citep{1992ARA&A..30...51B,2006MNRAS.365..555S}, and are thus either very long-lived or continuously regenerated, although both their origin and persistence remain topics of ongoing investigation. Current explanations invoke a tilt between the disk and triaxial dark matter halo, or a continuous infall of material with angular momentum misaligned with that of the disk \citep[e.g.,][]{2006ApJ...641L..33W,2006MNRAS.370....2S}. 

\subsection{The Gaia Reference Frame}

Gaia was launched on 19~December 2013. Over its 5-year program, progressively more accurate catalogs will be released as the continuous sky scanning increases the number of individual measurements per star, and simultaneously extends the temporal baseline. Details of the astrometric data processing are given by \citet{2012A&A...538A..78L}. As a result of on-board detection thresholding, Gaia will observe {\it all\/} star-like objects down to a completeness limit of $V\sim20$\,mag (more strictly, the astrometry integrates over a broad-band response designated~$G$). Out of its expected harvest of more than a billion objects, some 500\,000 or more quasars are expected to be observed and identified, mostly in the range $z=1.5-2.0$ \citep{2006MNRAS.367..879C,2012MmSAI..83..918M}. This will permit direct connection to an inertial reference system, with an accuracy estimated below.

Linking the Gaia catalog to the ICRS proceeds conceptually as follows: 
(a)~the observations are reduced to an internally consistent catalog of positions and proper motions, with arbitrary system orientation and spin;
(b)~positions of the optical counterparts of radio sources in the ICRF will be compared with their radio positions, to give the orientation vector $\bfvarepsilon$ of the optical catalog with respect to ICRF;
(c)~the (apparent) proper motions of quasars will be analysed to determine the quasi-inertial spin vector $\bfomega$ of the catalog with respect to the extragalactic frame. 
The final Gaia catalog then results from applying a correction corresponding to $-\bfvarepsilon$ to all positions, and by applying a correction corresponding to $-\bfomega$ to all proper motions. In practice, these steps will be incorporated within the iterative astrometric core processing \citep{2012A&A...538A..78L,2011ExA....31..215O}.

\section{The Gaia inertial frame in practice}

\subsection{Galactocentric acceleration}
\label{sec:galacceleration}

In the practical realization of a non-rotating inertial reference frame at the \muas\ level, the non-uniformity of the Galactic motion of the solar system barycenter is a manifestly non-negligible violation of inertiality. The principal observable effect is caused by the nearly constant (secular) acceleration of the barycenter with respect to the center of the Galaxy \citep{1995ESASP.379...99B,2003A&A...404..743K,2006AJ....131.1471K}. This acceleration causes the aberration term to change slowly with time, and therefore results in a pattern of secular aberration observable as a systematic vector field of the apparent proper motions of distant quasars. The effect has been observed as residuals in the VLBI reference frame \citep{2010MNRAS.407L..46T,2011A&A...529A..91T,2012A&A...544A.135X}, and has been identified as a contributing term in the orbital period change of the binary pulsar PSR~1913+16 \citep{1991ApJ...366..501D}.

The solar system's orbital velocity around the Galactic center, which we will adopt as $V_0=223$\kms\ (see below), causes an aberration effect of $V_0/c\sim2.5$\,arcmin; its absolute velocity with respect to a cosmological reference frame similarly causes the dipole anisotropy of the cosmic microwave background. All measured star and quasar positions are therefore shifted towards Galactic coordinates $l=90^\circ$, $b=0^\circ$. For an arbitrary point on the sky the size of the effect is 2.5\,arcmin~($\sin\eta$), where $\eta$ is the angular distance to the point $l=90^\circ$, $b=0^\circ$. 
Adopting Oort constants $A=14.82,B=-12.37$ (both in \kms\,kpc$^{-1}$) as derived from Hipparcos Cepheids \citep{1997MNRAS.291..683F} giving $\Omega_0=A-B=27.19$\kms\,kpc$^{-1}$, and a Galactocentric radius of the Sun of $R_0=8.2$\,kpc \citep{2008ApJ...689.1044G}, results in a circular velocity at $R_0$ of $V_0\equiv R_0\Omega_0=223$\kms, and a Galactic orbital period for the Sun of $P_{\rm rot}=2.26\times10^8$\,yr. 
The resulting Galactocentric acceleration of the barycenter has the value
\begin{equation}
\label{equ:galactocentric-acceleration}
a_{\rm Gal}\equiv \frac{V_0^2}{R_0}=2\times10^{-10}\,{\rm m}\,{\rm s}^{-2} = 6\times10^{-3}\,{\rm m\,s}^{-1}\,{\rm yr}^{-1}\ .
\end{equation}
This causes a change in (first-order) aberration of $a_{\rm Gal}/c\sim4$\muasyr, resulting in an apparent proper motion of a celestial object, towards the Galactic center, of 4\muasyr~($\sin\zeta$). This holds for all objects beyond about 200\,Mpc, and in particular for quasars, for which their intrinsic proper motions, caused by real transverse motions, are assumed negligible. A proper motion of 4\muasyr\ corresponds to a transverse velocity of $\sim$30\,000\kms\ at $z=0.3$ for $H_0=70$\kms\,Mpc$^{-1}$. Thus, all quasars will exhibit a distance-independent streaming motion towards the Galactic center. Within the Galaxy, on the other hand, the effect will be hidden in the local kinematics, e.g.,\ corresponding to $\sim0.2$\kms\ at 10\,kpc.

\subsection{Spin vector}

The spin vector, $\bfomega$, will be determined from the $\sim$500\,000 quasars, in the range $V=12-20$\,mag which will be observed by Gaia directly. Some of these, including large numbers from 2dF \citep{2004MNRAS.349.1397C} and SDSS \citep{2014A&A...563A..54P} will be known {\it a priori}. In any case, all will be detected on-board and therefore observed astrometrically and photometrically. Detailed studies \citep{2006MNRAS.367..879C} have shown that multi-parameter classification (based on colour indices, photometric variability, and negligible parallax and proper motion) will be able to identify a large fraction of those quasars previously unknown, at the same time excluding stars at some expense of completeness (an essential process given that quasars will represent only some 0.05\% of the observed objects).

For assessments of the accuracy of the link, the cumulative number density of quasars as function of magnitude was taken from \citet{1990ARA&A..28..437H}, and restricted to redshifts $z<2.2$. These authors already pointed out that the knowledge of the quasar luminosity function for $z<2.2$ `appeared to be quite secure'. This conclusion is broadly confirmed by the latest quasar compilation of \citet[][with sky distributions given by \citealt{2012MmSAI..83..918M}]{2010A&A...518A..10V}, complemented by a highly-simplified full-sky extrapolation of the 2dF \citep{2004MNRAS.349.1397C} and SDSS \citep{2014A&A...563A..54P} yields. At the same time, restriction to redshifts $z<2.2$ probably gives some underestimate of the final numbers expected to be available for the link; larger surface densities were estimated by \citet{2012MmSAI..83..918M} from an extrapolation of the highest densities found in \citet{2010A&A...518A..10V}. Our adopted, and probably conservative, numbers are given in Tables~\ref{tab:qso-link-10} and \ref{tab:qso-link-100}.

\subsection{Quasar source instabilities}
\label{sec:instabilities}

The positional/proper motion stability of individual quasars will be affected by: \\
(1)~macrolensing by intervening galaxies: this may cause apparent proper motions of several~\muasyr, but only if the impact parameter is close to the critical value (of the order of 1\,arcsec) where significant magnification occurs \citep{1996ApJ...473..610K}. The fraction of affected quasars is of the order of 1\% \citep{1996ApJ...457..228K}, and they usually have additional structure (multiple images and arcs) on scales that will be resolved by Gaia. For larger impact parameters, the proper motion of the single deflected image is smaller than the proper motion of the lensing galaxy, i.e.\ $\la 0.2$\muasyr\ for a lens at $z\sim0.1$; \\
(2)~gravitational lensing by stars in the Galaxy: some 1000 strong-lens quasars are expected to be diskovered by Gaia \citep{2006MNRAS.367..879C}, and excluded from the reference frame link. All quasars will be subject to weak lensing \citep{2002A&ARv..10..263C}, leading to random, variable displacements of $\sim$1\muas\ \citep{1998MNRAS.300..287S}. The typical effect on the mean proper motion over the Gaia lifetime will be $\la1$\muasyr; \\
(3)~photocentric motion: most of the quasar optical emission comes from a region of $\la1$\,pc, corresponding to $\la 200$\muas\ at 1\,Gpc. Assuming that the photocenter moves randomly within this region, a mean proper motion of $\la 50$\muasyr\ may result over the 5-year observation period. In a detailed study of the ultimate celestial reference for Gravity Probe~B, the superluminal quasar 3C~454.3 has a 7-yr proper motion limit of $<56$\muasyr. Photocentric motion is also induced by a variable nucleus combined with the much fainter, but much larger galaxy \citep[e.g.,][]{2011A&A...526A..25T}. This effect could reach some 100\muasyr, but extreme cases might be recognized by the correlation between position and brightness; \\
(4)~chromatic image displacement: although the Gaia telescopes are all-reflective, they are nevertheless not strictly achromatic. Asymmetric wavefront errors, such as coma, introduce image centroids that depend on wavelength, and hence on the object's spectral energy distribution. For the typical wavefront errors expected in the astrometric field, of $\sim$50\,nm rms, the centroid shift between early and late spectral types could reach several~mas. This systematic `chromaticity' effect can therefore be many times larger than the photon statistical uncertainty of the estimated image location. It is thus essential to have a very good calibration of the spectral energy distribution of each observed source, obtained on-board from the blue and red photometers \citep{2006MNRAS.367..290J}. Since quasar spectra potentially show strong emission lines at any wavelength depending on redshift, their chromaticity correction will be more problematic than for stars, and could generate spurious proper motion of instrumental origin of $\sim10$\muasyr. 

In summary, the most important instabilities are expected to be due to variable source structure and residual telescope chromaticity. The likely range of the combined effects for typical quasars may lie between 10--100\muasyr\ \citep[a value of 30\muasyr\ in each coordinate was inferred by][]{1997ApJ...485...87G}. These limits are used in the simulations described below, as summarized in Tables~\ref{tab:qso-link-10} and \ref{tab:qso-link-100} respectively.

\subsection{Condition equations}

The astrometric processing of the Gaia observations determines the positions, parallaxes and proper motions of stars and quasars in an internally consistent, but provisional reference frame \citep{2012A&A...538A..78L}. In this frame the quasars will have non-zero proper motions $(\mu_{l*},\,\mu_b)$ due to: 
(i)~the spin, $\bfomega$, of the provisional frame with respect to the cosmological reference frame;
(ii)~the apparent streaming motion caused by the acceleration, ${\bf a}$, of the solar system barycenter; and 
(iii)~observational errors and source instability. 
The first two effects are systematic while the third is assumed to be random and uncorrelated among the quasars. The spin vector should be determined simultaneously with the the acceleration vector in a single least-squares solution, using the apparent proper motions of all the quasars. For a quasar at Galactic coordinates $(l,~b)$ the condition equations for the Galactic components of $\bfomega$ and ${\bf a}$ are then:
\begin{eqnarray}
\mu_{l*} \equiv \mu_l\cos b 	&=&	{\bf q}' {\bfomega} + {\bf p}' {\bf a}\, c^{-1} + {\rm noise}\label{equ:cond1}\\
\mu_b 					&=&	{\bf p}' {\bfomega} + {\bf q}' {\bf a}\, c^{-1} + {\rm noise}\label{equ:cond2}
\end{eqnarray}
where $c$ is the speed of light, and ${\bf p} = (\sin l,\,\cos l,\,0)'$, ${\bf q} = (\sin b\cos l,\,\sin b\sin l,\,\cos b)'$ are unit vectors along $+l$, $+b$ tangent to the celestial sphere at the position of the quasar.

\subsubsection{Simulations}

Numerical simulations were made of the least-squares solution of $\bfomega$ and ${\bf a}$, with the following assumptions (see Tables~\ref{tab:qso-link-10} and \ref{tab:qso-link-100}). The available quasar numbers were randomly distributed over the sky, except in the Galactic plane $|b|<20^\circ$, where zero density was assumed. Only a fraction $P(V)$ of all the quasars is used; this approximates to the use of various photometric and astrometric criteria to reject possible stars \citep{2006MNRAS.367..879C}. Galactic coordinates were transformed to the ecliptic system, and the standard errors in $\mu_{\lambda*}$, $\mu_\beta$ were computed as a function of magnitude, and ecliptic latitude, $\beta$. A separate least-squares solution was made for each magnitude interval from 14--20, and one solution for the whole magnitude range. Only the covariance matrices are of interest; they were transformed back to the Galactic system, yielding the accuracy estimates in Tables~\ref{tab:qso-link-10} and \ref{tab:qso-link-100}.

To account for source instabilities, the quantity $\sigma_0$ (Sect.~\ref{sec:instabilities}) was added in quadrature to the formal proper motion uncertainties. For Table~\ref{tab:qso-link-10}, an optimistic value of $\sigma_0=10$\muasyr\ was assumed, while for Table~\ref{tab:qso-link-100} the assumption was a rather pessimistic $\sigma_0=100$\muasyr. Sub-$\mu$as accuracy in the spin components is nevertheless reached in both cases due to the large number of sources. The accuracy is slightly lower about $\omega_3$ (normal to the Galactic plane) than about the other two axes, due to the zone of avoidance. Comparable values have been considered by \citet{2012A&A...547A..59M}.

The solution for the acceleration ${\bf a}$ is practically orthogonal to that of $\bfomega$, and of equal accuracy when expressed in comparable units (${\bf a}/c$ has the dimension of proper motion, with 1\muasyr\ $\equiv4.606\times 10^{-11}$\,m\,s$^{-2}$). The Galactocentric acceleration of the solar system barycenter (Eqn.~\ref{equ:galactocentric-acceleration}) should be measurable at 5--10\% relative accuracy.

\subsubsection{Analytical solution}

The simple structure of the condition equations also allows for an analytical accuracy estimate by making some plausible statistical assumptions. If quasars of apparent magnitude $V$ are uniformly distributed on the celestial sphere, and the noise terms in the condition equations are uncorrelated with a standard deviation $\sigma_\mu$ that only depends on $V$, it is found that the determinations of the six unknowns $\omega_i$, $a_i\,c^{-1}$ (where $i=1,\,2,\,3$ for the Galactic axes) are approximately uncorrelated, each having a standard deviation given by
\begin{equation}\label{eqn:sd}
\sigma^2 \simeq \frac{3}{2} \left( \sum_V N(V)\  \sigma_\mu^{-2} (V) \right)^{-1} 
\end{equation} 
where $N(V)$ is the number of useful quasars per magnitude bin, and $\sigma_\mu$ combine the effects of observational errors and source stability.
A comparison with the preceding simulations, which used a more detailed model of the quasar distribution (e.g., assuming no useful quasars for $|b|<20^\circ$) along with inhomogeneous observational noise, shows that Eqn.~\ref{eqn:sd} is accurate within $\pm 20$\% for the same total number, an adequate agreement given the likely uncertainties related to source instabilities and in the actual number of useful quasars.

\subsection{Frame orientation}

Although unimportant for any kinematic interpretation, we can similarly estimate the accuracy of the Gaia reference frame orientation, $\bfvarepsilon$, with respect to the inertial frame. This will be established, consistent with the ICRF, by comparing radio sources positions in ICRF with those of their optical counterparts observed by Gaia.  The number of radio sources in ICRF2 is currently 3414 \citep{2009ITN....35....1M}, with positional uncertainties of $\ga$40\muas. We assume that half can be observed optically by Gaia, and that most will be faint ($V\sim19$) with positional accuracies of 100--200\muas. Eqn.~\ref{eqn:sd} then suggests that the Gaia frame orientation will be defined with an uncertainty of $\sim5-10$\muas\ in each component of $\bfvarepsilon$.

\section{Signature of the tilting disk}

The accuracy with which the final Gaia catalog represents an inertial frame is given by the uncertainties of $\bfvarepsilon$ and $\bfomega$. We have shown that the Gaia catalog will improve the accuracy of the optical materialization of ICRS by more than two orders of magnitude, allowing an examination of individual and bulk motions in the Galaxy's disk populations, with an enormous range of kinematic and dynamical applications. In the context of the Galaxy warp, for example, Gaia will extend detailed kinematic analyses to the probable disk edge, at $R\sim15$\,kpc, or some 7\,kpc from the Sun, where the warp induces a mean offset out of the plane of $\sim$1\,kpc.

Specifically, Gaia will permit the identification of large-scale disk torques due to the progressive collapse of matter as guided by the $\Lambda$CDM structure formation paradigm. For example, a bulk rotation of the disk with a characteristic rate of 2\,rad/Hubble time (30\muasyr) about an axis in the plane of the Galaxy \citep{2004ApJ...616...27B,2007MNRAS.380..657B} will significantly exceed the inertial reference frame residual rotation of some 0.2--0.5\muasyr.  If the disk and halo are misaligned \citep[e.g.,][]{2013MNRAS.434.2971D}, then Gaia should detect a disk rotation rate that depends on Galactocentric radius.

The practical detection of such bulk motions may be viewed as follows. If the stars have, in addition to their component of Galactic rotation (of about 5000\muasyr), an extra rotation of 30\muasyr\ about an axis in the Galactic plane, then the net effect is a rotation about an axis that is offset by $\arctan(30/5000)=0\ddeg4$ from the normal to the Galactic disk. Whether the plane of the disk can be determined that accurately from Gaia is not yet evident. However, there would also be a differential effect: assuming a flat rotation curve, the Galactic rotation varies from 10\,000 to 2500\muasyr\ between $R=4-12$\,kpc, so the offset would vary from $0\ddeg2-0\ddeg6$, and there would be a differential effect of similar magnitude when comparing stars at different Galactic radii. This would create an additional warp-like structure in the kinematics, identifiable independently of the quasars.  If the inner and outer disk are misaligned, then this will alter the radial dependance of this structure.

Might other effects mask these bulk motions? The small value of the CMB quadrupole seen by COBE, WMAP and Planck strongly constrains the net rotation of the Universe \citep{1985MNRAS.213..917B}, while the low amplitude of the large-angle CMB modes also constrains any large-scale bulk quasar motions. Thus, scalar and vector perturbations terms are not likely to be significant. Gravitational waves may introduce additional structure in the apparent quasar proper motions over the sky, over a wide range of frequencies from the inverse of the observation period up to the Hubble time, but composed primarily of second-order transverse vector spherical harmonics \citep{1997ApJ...485...87G,2004NewAR..48.1483J}.

\section{Conclusions}

Our simulations using realistic quasar counts show that an accuracy of better than 1\muasyr\ should be reached in all three inertial spin components of the Gaia reference frame, $\bfomega$, even assuming somewhat conservative numbers of quasars used for the link, and rather pessimistic assumptions on the effects of variable source structure.  At the same time, the Gaia reference frame orientation will be defined with respect to the ICRF with an uncertainty of $\sim5-10$\muas\ in each component of $\bfvarepsilon$, while the Galactocentric acceleration of the solar system barycenter will be measured at 5--10\% relative accuracy.

These tight constaints on the inertial spin will allow the interpretation of individual and bulk motions in the Galaxy disk populations within the framework of an inertial reference frame defined by distant quasars.

Bulk stellar motions in the direction of Galactic rotation will reflect the many known complexities of the Galaxy's disk and halo structure, and its (differential) rotational motion \citep[e.g.,][]{2007AJ....134..367M}. To this extent, a variety of effects will likely mask any time-dependent influence of any external (e.g., halo-driven) effects on the spin component~$\omega_3$.

Bulk rotational motions about axes in the plane of the Galaxy ($\omega_1,\omega_2$) will reflect tilting of the Galaxy disk, regardless of origin, combined with any warp-like motions. Assuming that the latter are important only somewhat outside the solar circle, and assuming that the disk interior to the solar cycle responds as a solid body, then torque-induced motions of order 2\,rad\,$H_0^{-1}\simeq30$\muasyr, will formally be significantly above the accuracy with which the spin components are constrained from quasar observations.

Detection of a time-dependent rotation of the angular momentum of the of the Galactic disk population would contribute to an undertanding of the dynamic effects of the dark halo on the disk (a basic test of Newtonian gravity) and will likely elucidate the dynamical history of the Milky Way. For example, the measurement of the halo rotation rate may change our interpretation of the Sagittarius stream, which appears to lie along the unstable intermediate axis orbit \citep{2010ApJ...714..229L,2013MNRAS.434.2971D}. In rapidly rotating halos, this intermediate axis orbit is stabilized \citep{1982ApJ...258..490H}. Halo figure rotation would also alter models of the dynamics of the Magellanic Clouds \citep{2010ApJ...721L..97B}.

\section{Acknowledgements}
We would like to thank the referee, Victor Debattista, for helpful comments. 


\begin{deluxetable}{ccccccccccccc}
\tablecaption{Residual spin of the Gaia reference frame, $\sigma(\omega_i)$, and Galactocentric acceleration of the solar system barycenter, $\sigma(a_i/c)$, estimated from a simulation of quasar observations. This table assumes a contribution of $\sigma_0=10$\muasyr\ from source instability.\label{tab:qso-link-10}}
\tabletypesize{\footnotesize}
\tablewidth{0pt}
\tablehead{
\colhead{$V$} && \colhead{$P$} & \colhead{$N_{\rm QSO}$} & \colhead{$\sigma_{\mu,\rm tot}$} 
 && \colhead{$\sigma(\omega_1)$} & \colhead{$\sigma(\omega_2)$} & \colhead{$\sigma(\omega_3)$}
 && \colhead{$\sigma(a_1/c)$} & \colhead{$\sigma(a_2/c)$} & \colhead{$\sigma(a_3/c)$} \\
(mag) && & & ($\mu$as~yr$^{-1}$) && \multicolumn{3}{c}{($\mu$as~yr$^{-1}$)} && \multicolumn{3}{c}{($\mu$as~yr$^{-1}$)} 
}
\startdata
\phm{00}$\le$\,\,\,15 && 1.0 & \phm{000\,0}40 & \phm{0}14 && 2.5\phm{0} & 2.5\phm{0} & 3.0\phm{0} && 2.5\phm{0} & 2.5\phm{0} & 3.0\phm{0} \\
15 -- 16 && 1.0 & \phm{000\,}230 & \phm{0}21 && 1.5\phm{0} & 1.5\phm{0} & 1.8\phm{0} && 1.5\phm{0} & 1.5\phm{0} & 1.8\phm{0} \\
16 -- 17 && 0.9 & \phm{00}1\,230 & \phm{0}30 && 0.93 & 0.93 & 1.14 && 0.93 & 0.93 & 1.14\\
17 -- 18 && 0.8 & \phm{0}11\,500 & \phm{0}45 && 0.46 & 0.46 & 0.57 && 0.46 & 0.46 & 0.57\\
18 -- 19 && 0.6 & \phm{0}60\,000 & \phm{0}74 && 0.33 & 0.33 & 0.41 && 0.33 & 0.33 & 0.41\\
19 -- 20 && 0.3 & \phm{0}97\,000 &            130 && 0.46 & 0.46 & 0.56 && 0.46 & 0.46 & 0.56\\
\cline{1-13}
\phm{00}$\le$\,\,\,20 &&       & \phm{}170\,000 &                 && 0.22 & 0.22 & 0.27 && 0.22 & 0.22 & 0.27\\
\enddata
\tablecomments{Columns contain, for each range of magnitude: 
$P$, assumed probability that a quasar is unambiguously recognized from photometric indices; 
$N_{\rm QSO}$, expected number of recognized quasars with $z<2.2$ and $|b|>20^\circ$; 
$\sigma_{\mu,\rm tot}$, mean standard errors in proper motion per object and coordinate, including a contribution of $\sigma_0=10$\muasyr\ from source instability; 
$\sigma(\omega_i)$, resulting precision of the spin components ($i=1$ towards the Galactic center, $i=2$ in the direction of Galactic rotation, $i=3$ towards the Galactic pole); 
$\sigma(a_i/c)$, the resulting precision of the acceleration of the solar system barycenter along the Galactic axes.}
\end{deluxetable}

\begin{deluxetable}{ccccccccccccc}
\tablecaption{As Table~\ref{tab:qso-link-10}, but with a contribution of $\sigma_0=100$\muasyr\ from source instability.
\label{tab:qso-link-100}}
\tabletypesize{\footnotesize}
\tablewidth{0pt}
\tablehead{
\colhead{$V$} && \colhead{$P$} & \colhead{$N_{\rm QSO}$} & \colhead{$\sigma_{\mu,\rm tot}$} 
 && \colhead{$\sigma(\omega_1)$} & \colhead{$\sigma(\omega_2)$} & \colhead{$\sigma(\omega_3)$}
 && \colhead{$\sigma(a_1/c)$} & \colhead{$\sigma(a_2/c)$} & \colhead{$\sigma(a_3/c)$} \\
(mag) && & & ($\mu$as~yr$^{-1}$) && \multicolumn{3}{c}{($\mu$as~yr$^{-1}$)} && \multicolumn{3}{c}{($\mu$as~yr$^{-1}$)} 
}
\startdata
\phm{00}$\le$\,\,\,15 && 1.0 & \phm{000\,0}40 & 100 && 17.5\phm{0} & 17.5\phm{0} & 21.4\phm{0} && 17.5\phm{0} & 17.5\phm{0} & 21.4\phm{0} \\
15 -- 16 && 1.0 & \phm{000\,}230 & 102 && \phm{0}7.4\phm{0} & \phm{0}7.4\phm{0} & \phm{0}9.0\phm{0} && \phm{0}7.4\phm{0} & \phm{0}7.4\phm{0} & \phm{0}9.0\phm{0} \\
16 -- 17 && 0.9 & \phm{00}1\,230 & 104 && \phm{0}3.3\phm{0} & \phm{0}3.3\phm{0} & \phm{0}4.0\phm{0} && \phm{0}3.3\phm{0} & \phm{0}3.3\phm{0} & \phm{0}4.0\phm{0} \\
17 -- 18 && 0.8 & \phm{0}11\,500 & 109 && \phm{0}1.12 & \phm{0}1.12 & \phm{0}1.37 && \phm{0}1.12 & \phm{0}1.12 & \phm{0}1.37\\
18 -- 19 && 0.6 & \phm{0}60\,000 & 124 && \phm{0}0.56 & \phm{0}0.56 & \phm{0}0.68 && \phm{0}0.56 & \phm{0}0.56 & \phm{0}0.68\\
19 -- 20 && 0.3 & \phm{0}97\,000 & 164 && \phm{0}0.58 & \phm{0}0.58 & \phm{0}0.71 && \phm{0}0.58 & \phm{0}0.58 & \phm{0}0.71\\
\cline{1-13}
\phm{00}$\le$\,\,\,20&&        & \phm{}170\,000 &          && \phm{0}0.38 & \phm{0}0.38 & \phm{0}0.46 && \phm{0}0.38 & \phm{0}0.38 & \phm{0}0.46\\
\enddata
\end{deluxetable}

\end{document}